\documentclass{sigchi}

\CopyrightYear{2019}
\setcopyright{rightsretained}
\isbn{}

\usepackage{balance}       
\usepackage{subfig}
\usepackage{graphicx}
\usepackage[T1]{fontenc}   
\usepackage{txfonts}
\usepackage{mathptmx}
\usepackage[pdflang={en-US},pdftex]{hyperref}
\usepackage{color}
\usepackage{booktabs}
\usepackage{textcomp}
\usepackage{cuted}
\usepackage{capt-of}
\usepackage{array}
\newcolumntype{P}[1]{>{\centering\arraybackslash}p{#1}}

\usepackage{microtype}        
\usepackage{ccicons}          

\usepackage{todonotes}

\newcommand{\systemname}{Sketchforme}

\def\plaintitle{\systemname{}: Composing Sketched Scenes from Text
Descriptions for Interactive Applications}
\def\plainauthor{Forrest Huang, John Canny}

\def\plainkeywords{interactive applications; natural language; sketching; generative models; deep learning; Transformer; interactive machine learning.}

\makeatletter
\def\url@leostyle{%
  \@ifundefined{selectfont}{
    \def\UrlFont{\sf}
  }{
    \def\UrlFont{\small\bf\ttfamily}
  }}
\makeatother
\def\pprw{8.5in}
\def\pprh{11in}

\definecolor{linkColor}{RGB}{6,125,233}
\hypersetup{%
  pdftitle={\plaintitle},
  pdfauthor={\plainauthor},
  pdfkeywords={\plainkeywords},
  pdfdisplaydoctitle=true, 
  bookmarksnumbered,
  pdfstartview={FitH},
  colorlinks,
  citecolor=black,
  filecolor=black,
  linkcolor=black,
  urlcolor=linkColor,
  breaklinks=true,
  hypertexnames=false
}

\renewcommand{\comment}[1]{}  

\begin{document}

\title{\plaintitle}



\title{\plaintitle}

\numberofauthors{2}
\author{%
  \alignauthor{Forrest Huang\\
    \affaddr{University of California, Berkeley}\\
    \affaddr{Berkeley, U.S.A.}\\
    \email{forrest\_huang@berkeley.edu}}\\
  \alignauthor{John F. Canny\\
    \affaddr{University of California, Berkeley}\\
    \affaddr{Berkeley, U.S.A.}\\
    \email{canny@berkeley.edu}}\\
}

\maketitle
\begin{strip}\centering
\includegraphics[width=\textwidth]{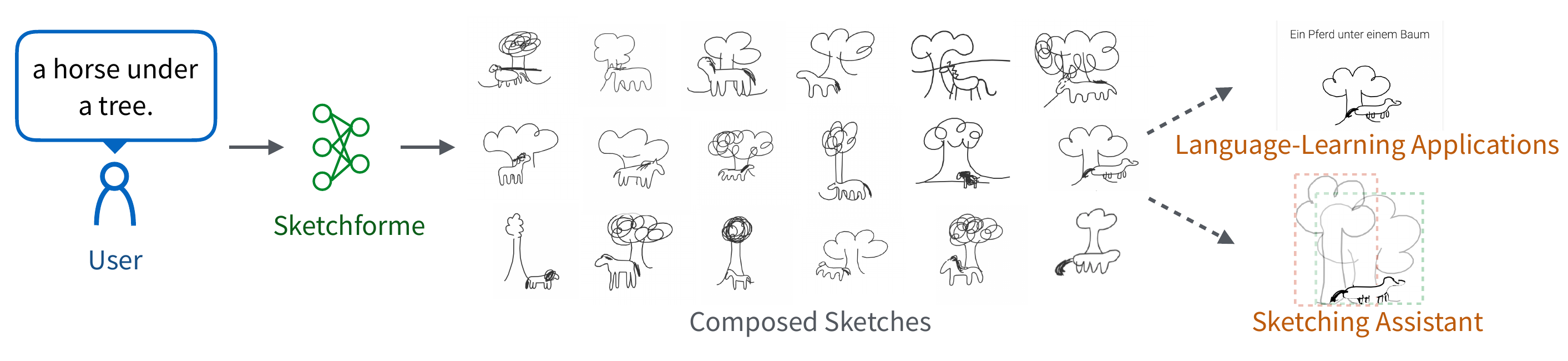}
\captionof{figure}{\systemname{} synthesizes sketched scenes corresponding to users' text descriptions to support interactive applications.
\label{fig:teaser}}
\end{strip}

\begin{abstract}
Sketching and natural languages are effective communication media for interactive applications. We introduce \systemname{}, the first neural-network-based system that can generate sketches based on text descriptions specified by users. \systemname{} is capable of gaining high-level and low-level understanding of multi-object sketched scenes without being trained on sketched scene datasets annotated with text descriptions. The sketches composed by \systemname{} are expressive and realistic: we show in our user study that these sketches convey descriptions better than human-generated sketches in multiple cases, and 36.5\% of those sketches are considered to be human-generated. We develop multiple interactive applications using these generated sketches, and show that \systemname{} can significantly improve language learning applications and support intelligent language-based sketching assistants.
\end{abstract}




\maketitle

\section{Introduction}
 
Sketching is a natural and effective way for people to communicate artistic and functional ideas. Sketches are \emph{abstract} drawings widely used by designers, engineers and educators as a thinking tool to materialize their vision while discarding unnecessary details. Sketching is also a popular form of artistic expression among amateur and professional artists. With the pervasive use of sketches across diverse fields, researchers in the HCI and graphics communities developed sketch-based interactive tools to enable intuitive and rich user experiences, such as assisted sketching systems \cite{inria, shadowdraw}, design tools for prototyping \cite{landay}, and animation authoring tools \cite{k-sketch}.

While using sketches as an interactive medium poses numerous benefits, producing meaningful and delightful sketches can be challenging for users and typically requires years of education and practice. This is partially due to the array of skills one needs to master for sketching: finding the appropriate abstraction level desired for the sketch, composing such abstraction with visual representation involving individual objects, and conveying these objects via precise motor operations with drawing media. Researchers have developed computational methods to assist users in creating sketches, but these methods typically rely on rigid modeling of the pixel/stroke-level visual details without semantic understandings of the sketched content. These restrict the use-cases of these approaches to improving sketching mechanics of user-created sketches.

Recent advances in neural-network-based generative models drastically increased machines' ability to generate convincing graphical content, including sketches, from high-level concepts. The Sketch-RNN model \cite{sketch-rnn} demonstrates recurrent neural networks (RNNs) trained on crowd-sourced data can understand and generate original sketches from various concept classes. Using these techniques, we introduce \emph{\systemname{}}, the first system that is capable of synthesizing sketches conditioned on natural language descriptions. 

The contribution of this paper is two-fold. First, we contribute \emph{\systemname{}}, the system that uses a novel two-step neural method for generating sketched scenes from text descriptions. \systemname{} first uses its \emph{Scene Composer}, a neural network that learned \emph{high-level composition principles} from datasets of human-annotated natural images that contain text captions, bounding boxes of individual objects and class information, to generate scenes composition layouts. \systemname{} then uses its \emph{Object sketcher}, a neural network that learned \emph{low-level sketching mechanics} to generate sketches adhering to the objects' aspect ratios in the composition. \systemname{} composes these generated objects of certain aspect ratios into meaningful sketched scenes. 

Second, we contribute and evaluate several applications, including a sketch-based language learning system and an intelligent sketching assistant, to exemplify the importance of \systemname{} in empowering novel sketch-based applications in Section \ref{sec:applications}.  In these applications, \systemname{} creates new interactions and user experiences with the interplay between language and sketches. We envision the connection that \systemname{} draws between language and sketching will enable further engaging and natural human-computer interactions, and open up new avenues for self-expression by users. 


\section{Related Work}

\subsection{Assisted sketching tools and tutorials}
Prior works have augmented the creative process of sketching with automatically-generated and crowd-sourced drawing guidance. ShadowDraw \cite{shadowdraw} and EZ-sketching \cite{ez-sketching} used edge images traced from natural images to suggest realistic sketch strokes to users. The Drawing Assistant \cite{inria} extracts geometric structure guides to help users construct accurate drawings. PortraitSketch \cite{potrait-sketch} provides sketching assistance specifically on facial sketches by adjusting geometry and stroke parameters. Researchers also developed crowd-sourced web applications to provide real-time feedback for users to correct and improve sketched strokes \cite{crowd}.

In addition to assisted sketching tools, researchers also developed sketch tutorial tools to improve users' sketching proficiency. How2Sketch \cite{how2sketch} automatically generates multi-step tutorials of sketching 3D objects. Sketch-sketch revolution \cite{sketch-sketch} provides first-hand experiences created by sketch experts for novice sketchers. 

While these methods help users create refined sketches, none of them can synthesize sketches from semantic descriptions as \systemname{}'s sketch generation process.

\subsection{Neural-network based sketch generation model}
\systemname{} builds upon the Sketch-RNN model, the first neural-network based sketch generation model \cite{sketch-rnn}. Sketch-RNN consists of a sequence encoder-decoder model that can unconditionally generate stroke-based sketches based on object classes, and conditionally reconstruct sketches based on users' input sketches. \systemname{} extends Sketch-RNN's model for sketching individual objects to support conditional sketch generation based on aspect ratios in the composition layouts.

\subsection{Neural-network-based text-to-image synthesis}
Generating graphical content from text description is a popular ongoing research problem. Recent work on Generative Adversarial Networks (GANs) \cite{gan-cls, stackgan} shows promising results in generating realistic images from text descriptions. GAN-CLS \cite{gan-cls} augments the GAN architecture to consider text descriptions, and subsequently generate images based on users' text input. Extending on these works, \cite{inferring-cgan} introduces multiple components to first synthesize composition and outlines from a text description, and subsequently generate images from these composition and outlines. This is similar to \systemname's multi-step approach to generate complete sketch-scenes from natural language except in the domain of natural images.

\subsection{Neural Style Transfer}
Several prior works in the Computer Vision community focus on the research problem of transferring styles between visual content. These prior works explore image stylization by matching statistics of feature maps (i.e. filters) of pre-trained models and with generative adversarial networks \cite{cyclegan, facial}. One possible approach for sketch generation that arises from these techniques is to stylize synthetic images generated based on text descriptions. However, this approach likely results in realistic, detailed sketch-style images which contain distracting artifacts. \systemname{} focuses on synthesizing \emph{abstract sketched scenes from scratch} that capture fundamental ideas from messages communicated by the scenes.

\section{System Description}
\label{sec:methods}
\begin{figure}[h]
\begin{center}
\includegraphics[width=1.0\linewidth]{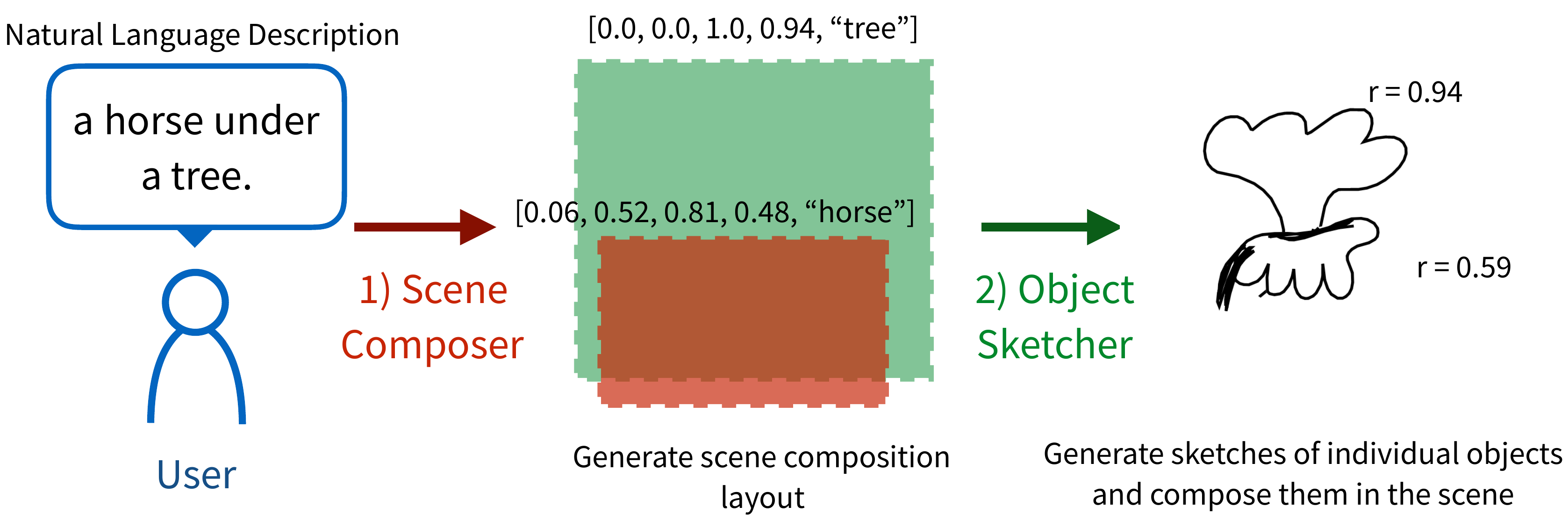}
\end{center}
\caption{Overall system architecture of \systemname{} which consists of two steps in its sketch generation process.}
\label{fig:archi}
\end{figure}

To support applications that affords sketch and natural-language based interactions, we developed \systemname{}, the system that provides the core capability of synthesizing sketched scenes from natural language descriptions. \systemname{} implements a two-step approach to generate a complete scene from text descriptions as illustrated in Figure \ref{fig:archi}. In the first step, \systemname{} uses its \emph{Scene Composer} to synthesize composition layouts represented by bounding boxes of individual objects. These bounding boxes dictate the location, size and aspect ratio of the objects in the scene. \systemname{}'s \emph{Object Sketcher} then uses this information at the second step of the generation process to generate specific sketch strokes of these objects in their corresponding bounding boxes. These steps are reflective of humans' sketching process of scenes suggested by sketching tutorials, where the overall composition of the scene is drafted before filling in details that characterize each object \cite{dummies}.

By taking this two-step approach, \systemname{} is able to obtain both high-level scene understanding and knowledge of the relation between individual objects. This enables a multitude of applications that require such understanding. Moreover, this approach overcomes the difficulty for end-to-end sketch generation methods to capture global structures of sequential inputs \cite{sketch-rnn}. End-to-end scene sketch generation also requires datasets of dedicated sketch-caption pairs that is difficult for crowd-workers to create \cite{sketchy-scenes} and will be prohibitively large in scale due to the combinatorial explosion of objects in the scenes.

\subsection{Scene Composer: Generating composition layouts}
\label{sec:cgan}
To generate composition layouts of scenes, we first model composition layouts as a sequence of $n$ objects, such that each object generated by the network is represented with 8 values: $$b_t = [x_t, y_t, w_t, h_t, l_t, \textrm{box}_t, \textrm{start}_t , \textrm{end}_t], t \in [1, n]$$ The first four values are fundamental data that describes bounding boxes of objects in the scene: x-position, y-position, width, height, and the class label. The last three values are boolean flags used as extra `tokens' to mark the actual boxes, the beginning of sequences and the end of sequences.

Using this sequential encoding of scenes, we designed a Transformer-based Mixture Density Network as our \emph{Scene Composer} to generate realistic composition layouts. Transformer Networks \cite{transformer} are state-of-the-art neural networks for sequence-to-sequence modeling tasks, such as machine translation and question answering. We use the Transformer to perform a novel task: generating a sequence of objects from a text description $c$, a sequence of words. As multiple scenes can correspond to the same text descriptions, we feed the outputs of the Transformer Network into Gaussian Mixture Models (GMMs) to model the variation of scenes forming a Mixture Density Network \cite{bishop}.

The generation process of the composition layouts involves taking the previous bounding box $b_{t-1}$ (or the start token) as an input and generating the current box $b_t$. At each time-step, the Transformer model generates an output $t_t$ conditioned on the text input $c$ and previously generated boxes $b_{1...t-1}$ using self-attention and cross-attention mechanisms built into the architecture. This process is repeated for multiple bounding boxes until an end token is generated:

\begin{equation}
    t_t = \text{Transformer}([b_{1...t-1}; c])
\end{equation}
$t_t$ is then projected to the appropriate dimensionality to parametrize the GMM models with various projection layers $W_{xy}$ and $W_{wh}$ to model $P(x_t, y_t)$, the distribution of the bounding boxes' positions, and $P(w_t, h_t)$, the distribution of the bounding boxes' sizes. With these distributions, \systemname{} can generate bounding boxes $[x_t, y_t, w_t, h_t]$ by sampling from these distributions in Equations \ref{eq:gmm1} and \ref{eq:gmm2}. The GMM models use the projected values as mean and co-variance for mixtures of multi-variate Gaussian distributions $M$. These parameters are passed through appropriate activation functions (Sigmoid, $\exp$ and $\tanh$) to comply with the required range of the parameters.

\begin{multline}
\label{eq:gmm1}
P(x_t, y_t) = \sum_{i \in M} \Pi_{1, i} \mathcal{N}(x_t, y_t | \mu_1, \Sigma_1),\\
[\Pi_1, \mu_1, \Sigma_1] = a(W_{xy}(t_t))
\end{multline}

\begin{multline}
\label{eq:gmm2}
P(w_t, h_t) = \sum_{i \in M} \Pi_{2, i} \mathcal{N}(w_t, h_t |\mu_2, \Sigma_2),\\
[\Pi_2, \mu_2, \Sigma_2] = a(W_{wh}([x_t; y_t; t_t]))
\end{multline}

While $P(x_t, y_t)$ is modeled only from the first projection layer $W_{xy}$, we consider $P(w_t, h_t)$ to be conditioned on the width and height of the boxes with the position of the boxes similar to \cite{inferring-cgan}. To introduce this condition, we concatenate the $t_t$ and the values of $[x_t, y_t]$ to the second projection layer as described in Equation \ref{eq:gmm2}. The probabilities of the current boxes are generated using a softmax-activated third projection layer $W_{c}$ from the Transformer output:

\begin{equation}
P(\text{box}_t, \text{start}_t, \text{end}_t) = \text{softmax}(W_{c}t_t)
\end{equation}

\begin{figure}[h]
\begin{center}
\includegraphics[width=0.75\linewidth]{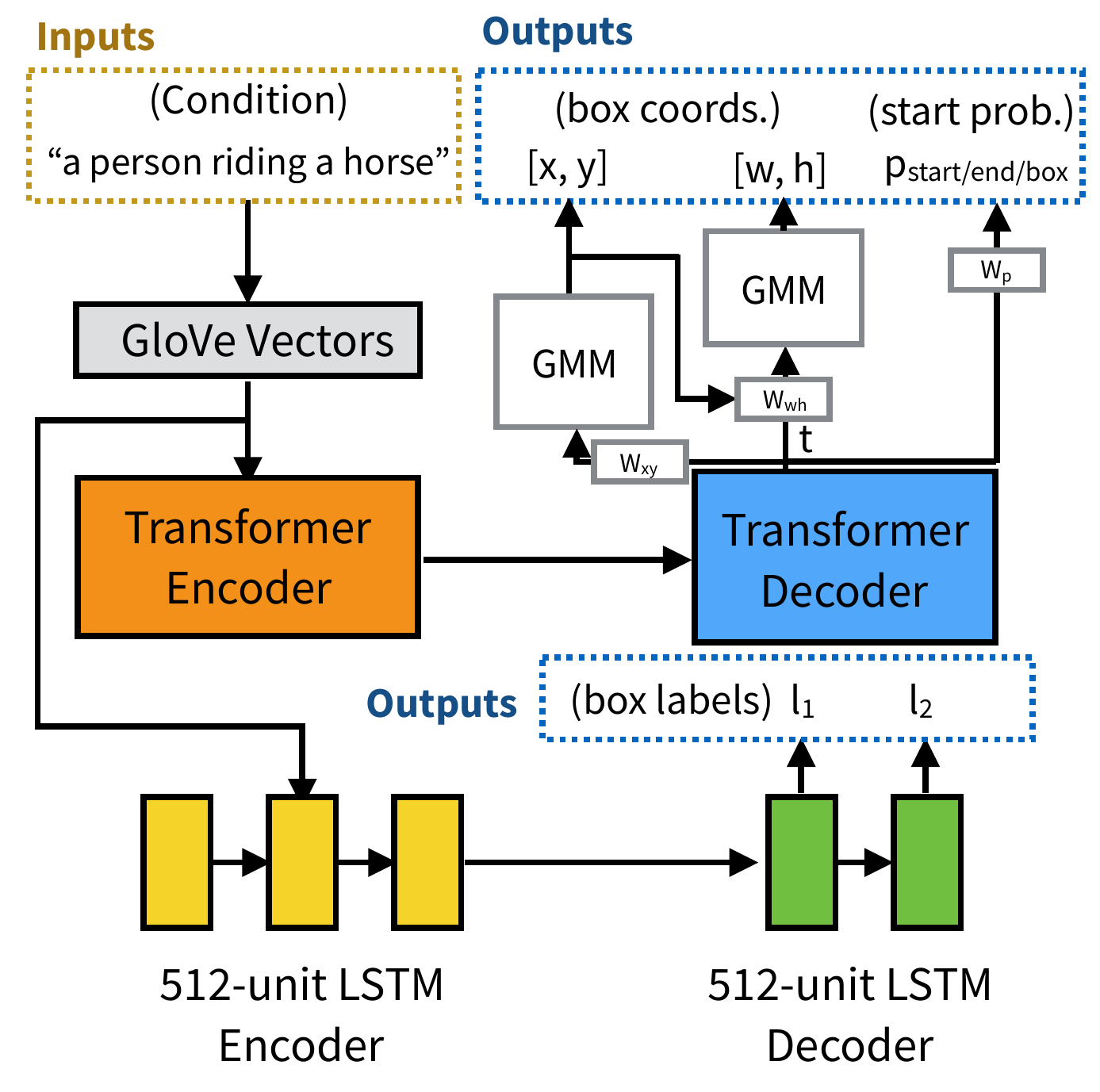}
\end{center}
\caption{Model architecture of the Scene Composer used in the first step of \systemname{}'s sketch generation process.}
\label{fig:gan_architecture}
\end{figure}

In addition, \systemname{} separately uses an LSTM to generate additional class label vectors $l_t$ because the class labels given certain descriptions are assumed to not vary across examples. The full architecture of the Scene Composer is shown in Figure \ref{fig:gan_architecture}.

\subsection{Object Sketcher: Generating individual sketches}
\label{sec:sketch-rnn}
After obtaining scene layouts from the Scene composer, we designed a modified version of Sketch-RNN to generate individual objects in \systemname{} according to the layouts. We adopt the decoder-only Sketch-RNN that is capable of generating sketches of only individual objects as a sequence of individual strokes. Sketch-RNN's sequential generation process involves generating the current stroke based on the previous strokes in the sketched object commonly used in sequence modeling tasks. Sketch-RNN also uses a GMM to model variation of sketch strokes.

While the decoder-only Sketch-RNN generates realistic sketches of individual objects in certain concept classes, the aspect ratio of the output sketches generated by the original Sketch-RNN model cannot be constrained. Hence, sketches generated by the original Sketch-RNN model generated sketches may be unfit for assembling into scene sketches guided by the layout generated by the Scene composer. Further, naive direct resizing of the sketches can produce sketches of unsatisfactory quality for complex scenes. 

\begin{figure}[h]
\begin{center}
\includegraphics[width=0.75\linewidth]{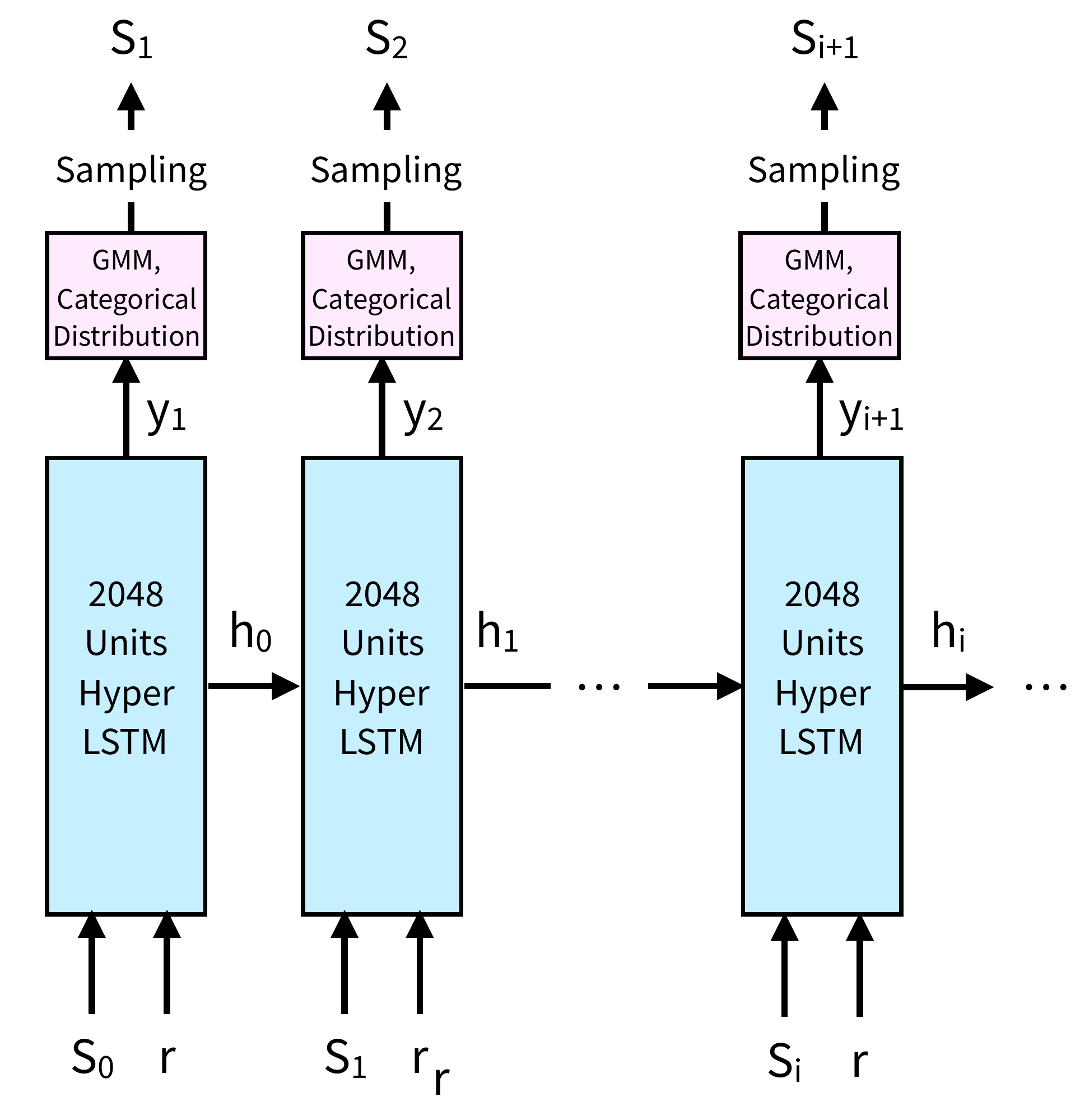}
\end{center}
\caption{Model architecture of the extended Sketch-RNN in \systemname's Object Sketcher.}
\label{fig:extended_rnn}
\end{figure}

We modified Sketch-RNN as the \emph{Object Sketcher} that factors in the aspect ratios of sketches when generating individual objects. To incorporate this knowledge in the Sketch-RNN model, we compute the aspect ratio of the training data and concatenate the aspect ratio $r = \frac{\Delta y}{\Delta x}$ of the sketch with the previous input stroke in the sketch generation process of our modified Sketch-RNN as shown in Figure \ref{fig:extended_rnn}. The new formulation and output of the Sketch-RNN at stroke-sequence $t$ is:

\begin{equation}
\label{eqn:extended_sketchrnn}[h_t; c_t] = LSTM([S_{t-1}; h_{t-1}; r]), y_t = Wh_t + b_t
\end{equation}

Since each Sketch-RNN model only handles a single object class, we train multiple sketching models and use the appropriate model based on the class label in the layouts generated by the Scene Composer for assembling the final sketched scene.

\section{Model Training and Data Sources}
\label{sec:training}
\systemname's Scene Composer and Object Sketcher are trained on different datasets that encapsulate visual-scene-level knowledge and sketching knowledge separately. This relaxes the requirement for \systemname{} to be trained on natural language annotated datasets of sketched scenes that provides highly varied scenes corresponding to realistic scene-caption pairs. 

We trained the Scene Composer using the Visual Genome dataset \cite{visual-genome} which contains natural language region descriptions and object relations of natural images to demonstrate its flexibility in utilizing various types of scene-layout datasets. Object relations in the dataset each contains a `subject' (e.g., `person'), an `object' (e.g., `on'), and a `predicate' (e.g., `car') represented by class labels and bounding boxes of the participating objects in the image. Natural language region descriptions are represented by bounding boxes of the regions and description texts that correspond to the regions. We reconcile these two types of information using region graphs in the dataset that pair these two types of data. With the paired data of natural language descriptions and relations, we train the Scene Composer to generate composition layouts. We selected relations that contain subsets of the 100 most commonly used object classes and 70 predicates in the dataset. This dataset of selected object classes and predicates contains 101,968 instances. We split this dataset in the scheme of: 70\% training set, 10\% validation set and 20\% testing set. 

The Object Sketcher is trained with the Quick, Draw! \cite{quickdraw} dataset that consists of 70,000 training sketches, 2,500 validation sketches and 2,500 testing sketches for each of the 345 object categories in the dataset. As mentioned in Section \ref{sec:methods}, we preprocess the data by computing the aspect ratio of each sketch as inputs to the Object Sketcher in addition to the original stroke data.

Using these data sources, we train multiple neural networks of various configurations and loss functions in \systemname. The LSTM architectures in the Scene Composer for generating composition layouts is stacked with 2 hidden-layers of size 512. Similarly, the Transformer Network has the configuration $(d_{model}, N_{layers}) = (512, 6)$.

The Scene Composer is trained by minimizing both the negative log likelihood of the position and size data:

\begin{equation}
L_{xy} = -\sum\limits_{i=1}^{n} \log(P(x_i, y_i)) 
\end{equation}
\begin{equation}
L_{wh} = -\sum\limits_{i=1}^{n} \log(P(w_i, h_i)) 
\end{equation}

and cross-entropy loss for categorical outputs $L_{P}$.

\begin{multline}
L_{p} = - (\sum\limits_{t=1}^{n} P(\text{box}_t) \log(\text{box}_t) + P(\text{start}_t) \log(P(\text{start}_t)) \\ + P(\text{end}_t) \log(P(\text{end}_t)))
\end{multline}

For generating the class labels, note that in our network each $l_t$ is represented as a 100-dimension vector, with each value $l_{i, t}$ corresponding to the output probability of the class. $L_{\text{class}}$ is thus computed as:
\begin{equation}
L_{\text{class}} = -\sum\limits_{t=1}^{n} \sum\limits_{i=1}^{100} l_{i, t} \log(l_{i, t}) \end{equation}

We combine these multiple losses with weight hyper-parameters to obtain a general training objective $L_{SC}$ for the Scene Composer:
\begin{equation}
L_{SC} = \lambda_1 L_{xy} + \lambda_2 L_{wh} + \lambda_3 L_{p} + \lambda_4 L_{\text{class}}
\end{equation}

The initial learning rate for the model is $1 \times 10^{-5}$. We use $\lambda_1 = 1.0, \lambda_2 = 1.0, \lambda_3 = 1 \times 10^{-5}, \lambda_4 = 1 \times 10^{-3}$. We use the Adam Optimizer with $\beta_1 = 0.9, \beta_2 = 0.999$ to minimize the loss function. We use 5 mixtures in each of the GMMs. We chose these hyper-parameters based on empirical experiments.

The Object Sketcher uses an HyperLSTM cell \cite{hyperlstm} of size 2048 for the modified Sketch-RNN model. The loss function of the Sketch-RNN model is identical to the reconstruction loss $L_R$ in the original Sketch-RNN model to maximize the log-likelihood of the generated probability distribution of the stroke data $S_t$ at each step $t$. The model is trained with an initial learning rate of $0.0001$ and gradient clipping of 1.0. 

\section{Experiments and Results}
Central to evaluating \systemname{}'s success is assessing its effectiveness in generating realistic and relevant sketches and layouts from text descriptions. We evaluated the data generated by \systemname{} at each step of the generation process qualitatively and quantitatively to demonstrate its effectiveness of generating sketched scenes. We further conducted a user study on the overall utility of the generated sketches to explore their potential in supporting real-world applications.

\subsection{Composition layouts generation}
The composition layouts generated by the Scene Composer in the first step of \systemname{} are represented as bounding boxes of individual objects in the scene. While the GMM in the Scene Composer already directly maximizes the log likelihood of the data, we can evaluate the performance of the model by visualizing and comparing heat-maps created by super-positioning instances of real data and generated data. 

Because \systemname{} considers the text input when generating the composition layouts, we should only compare the generated bounding boxes with bounding boxes from the dataset that is semantically similar to the text input. We obtain semantically similar ground-truth compositions by filtering the subjects, objects, and predicates based on the descriptions. For instance, the composition layouts generated from `a person riding a horse.' are compared with all actual compositions with a `person' subject, predicate that is related to riding such as 'on', 'on top of' etc. and `horse' subject.

\begin{figure}[ht]
  \centering
  \includegraphics[width=\linewidth]{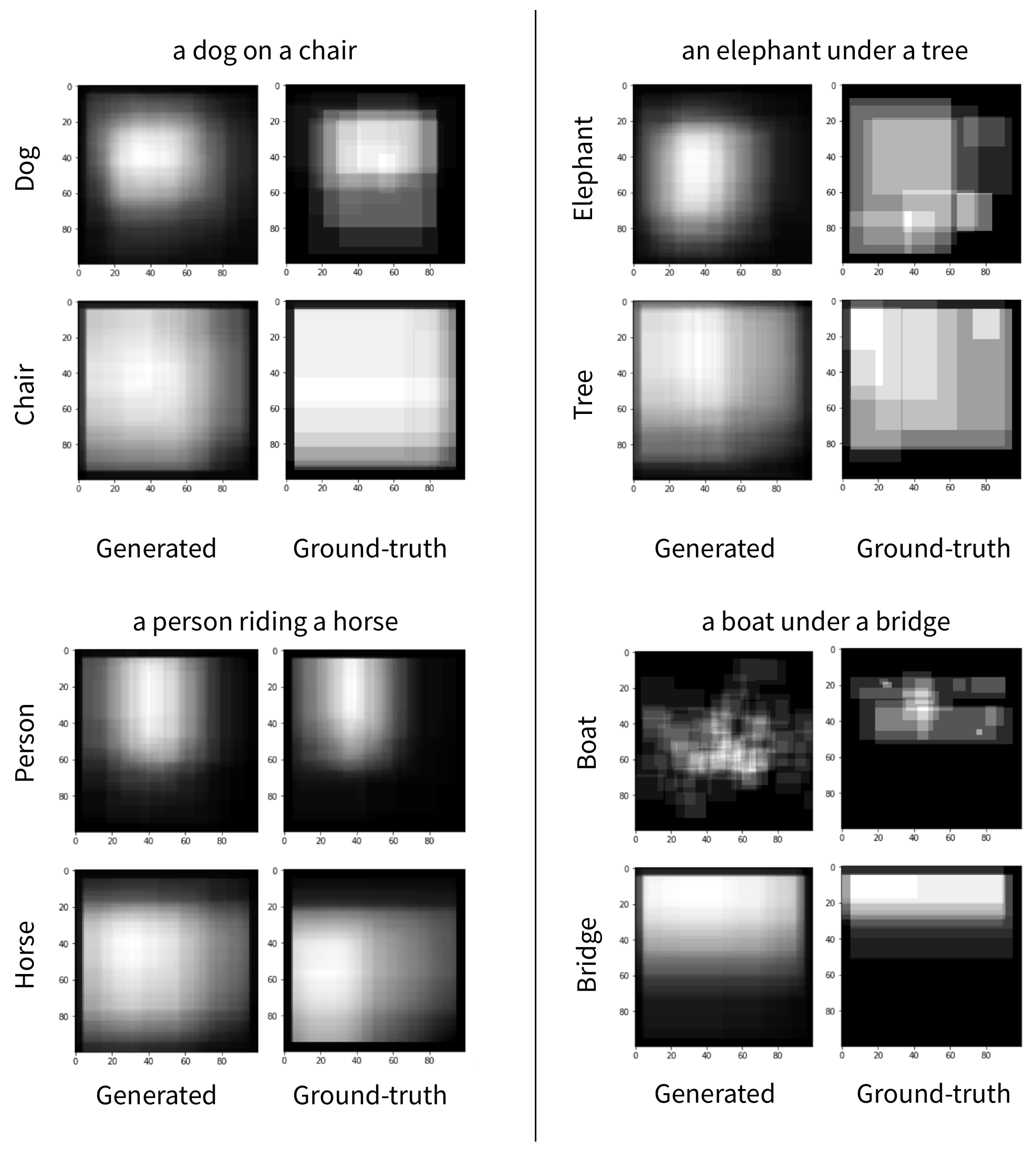}
  \caption{Heat-maps generated by super-positioning Generated/Visual Genome (ground-truth) data. Each horizontal pair of heat-maps corresponds to an object under a description.}
  \label{fig:heatmap}
\end{figure}

Heat-maps in Figure \ref{fig:heatmap} shows the distributions of \systemname{}'s synthetic bounding boxes and ground-truth bounding boxes from the dataset. From these heat-maps, we can obtain a holistic view on the generation performance of the model by visually evaluating the similarity between the heat-maps. We observe similar distributions between the actual relations and the generated composition layouts across all descriptions that correspond to the composition layouts. 

We can further approximate an overlap metric between the distributions using a Monte-Carlo simulation to obtain a quantitative metric of the model's performance. To estimate the overlap between the generated data distribution and the dataset's distribution, we generated 100 composition layouts for each prompt, and randomly sampled 1000 data points within each bounding boxes in these layouts. We estimate the overlap between the distributions by counting the number of data points that lie within the intersections between any generated and ground-truth bounding boxes. We compare \systemname{}'s performance with both a heuristic-based bounding box generator and a naive random bounding box generator. The heuristic-based bounding box generator only generates the second bounding boxes below the first bounding boxes for prompts with the 'above'-related predicates, and vice versa. The random bounding-box generator samples random values that describe the bounding boxes from uniform distributions serving as a naive baseline. Table \ref{table:jsdivergence} shows the percentage of the 1000 data points that lie in the intersections. The overlap between real data and \systemname{}-synthesized data is higher than both of that between the heuristic-based generator and the random generator by a large margin, which confirms our qualitative visual inspection of the heat-maps.

\begin{table}[ht]
\begin{center}
\begin{tabular}{P{2.4cm}|P{1.7cm}P{1.2cm}P{1.2cm}}
Description & \textbf{\systemname{}} & Heuristics & Random\\
\hline
a dog on a chair & \textbf{89.1\%} & 64.4\% & 61.6\% \\
\hline
an elephant under a tree           & \textbf{68.4\%} & 40.3\% & 30.6\% \\
\hline
a person riding a horse           & \textbf{94.0\%} & 57.7\% & 51.5\%  \\
\hline
a boat under a bridge     & \textbf{31.8\%} & 15.0\% & 6.85\% \\
\end{tabular}
\end{center}

\caption{Overlap metric from Monte-Carlo simulations for each description between real data and generated/heuristics-generated/random data.}
\label{table:jsdivergence}
\end{table}

\subsection{Generating individual object sketches at various aspect ratios}
\begin{figure}[h]
\begin{center}
\includegraphics[width=0.9\linewidth]{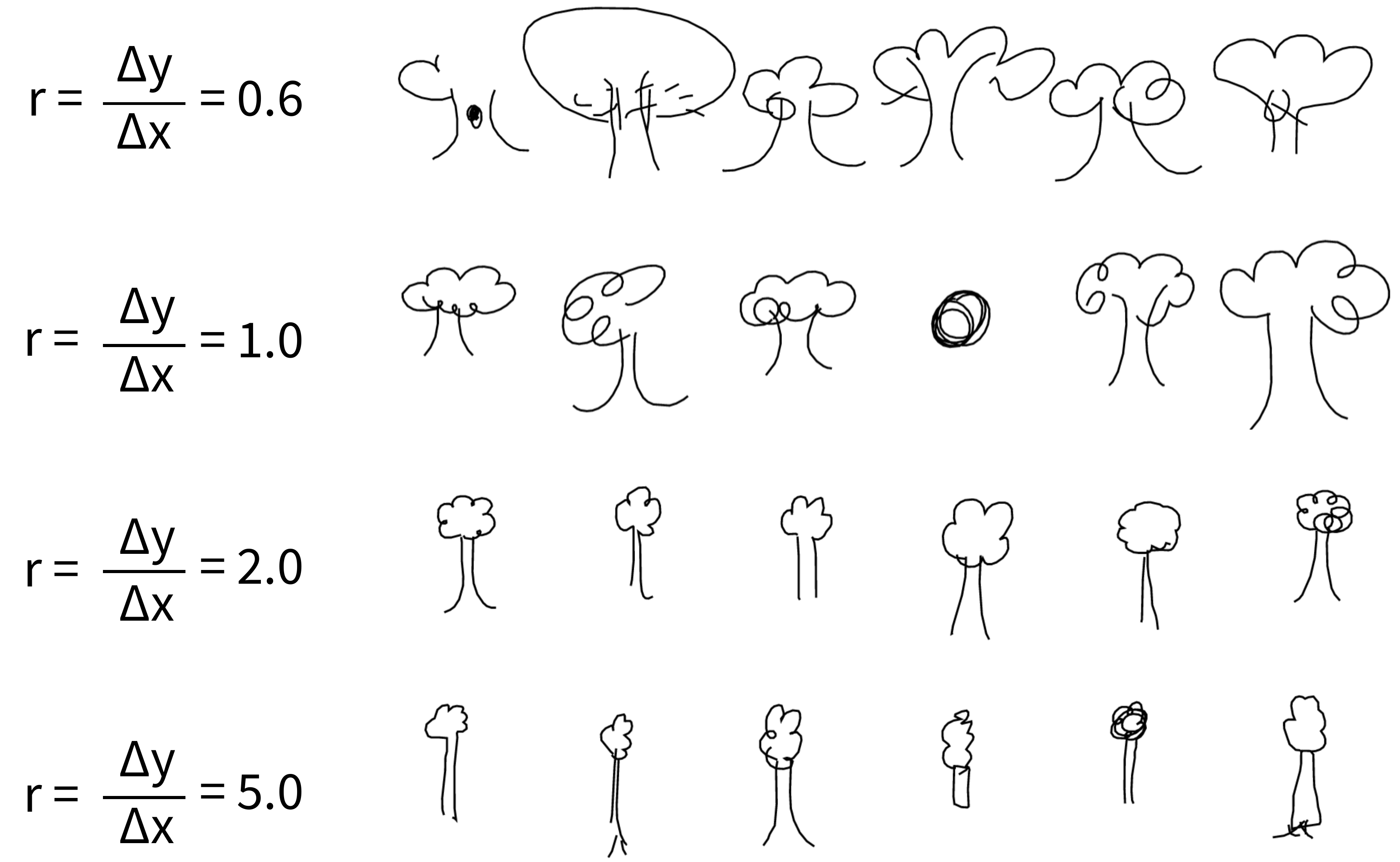}
\end{center}
\caption{Generated sketches of trees with various aspect ratios by the extended Sketch-RNN model in \systemname.}
\label{fig:aspect_ratios}
\end{figure}

The main addition of \systemname{} to the original Sketch-RNN model is an additional input that allows the Object Sketcher to generate sketches based on target aspect ratios ($r = \frac{\Delta y}{\Delta x}$) of completed sketches. We evaluate this approach by generating sketches of various aspect ratios. The Object Sketcher is able to adhere to input aspect ratios and generate individual object sketches coherent to the ratios. As shown in Figure \ref{fig:aspect_ratios}, trees generated with ratio $r = 1.0$ can be perceived as significantly shorter than trees with $r = 2.0$.

\subsection{Complete scene sketches} 
\begin{figure}[ht]
  \centering
  \includegraphics[width=1.0\linewidth]{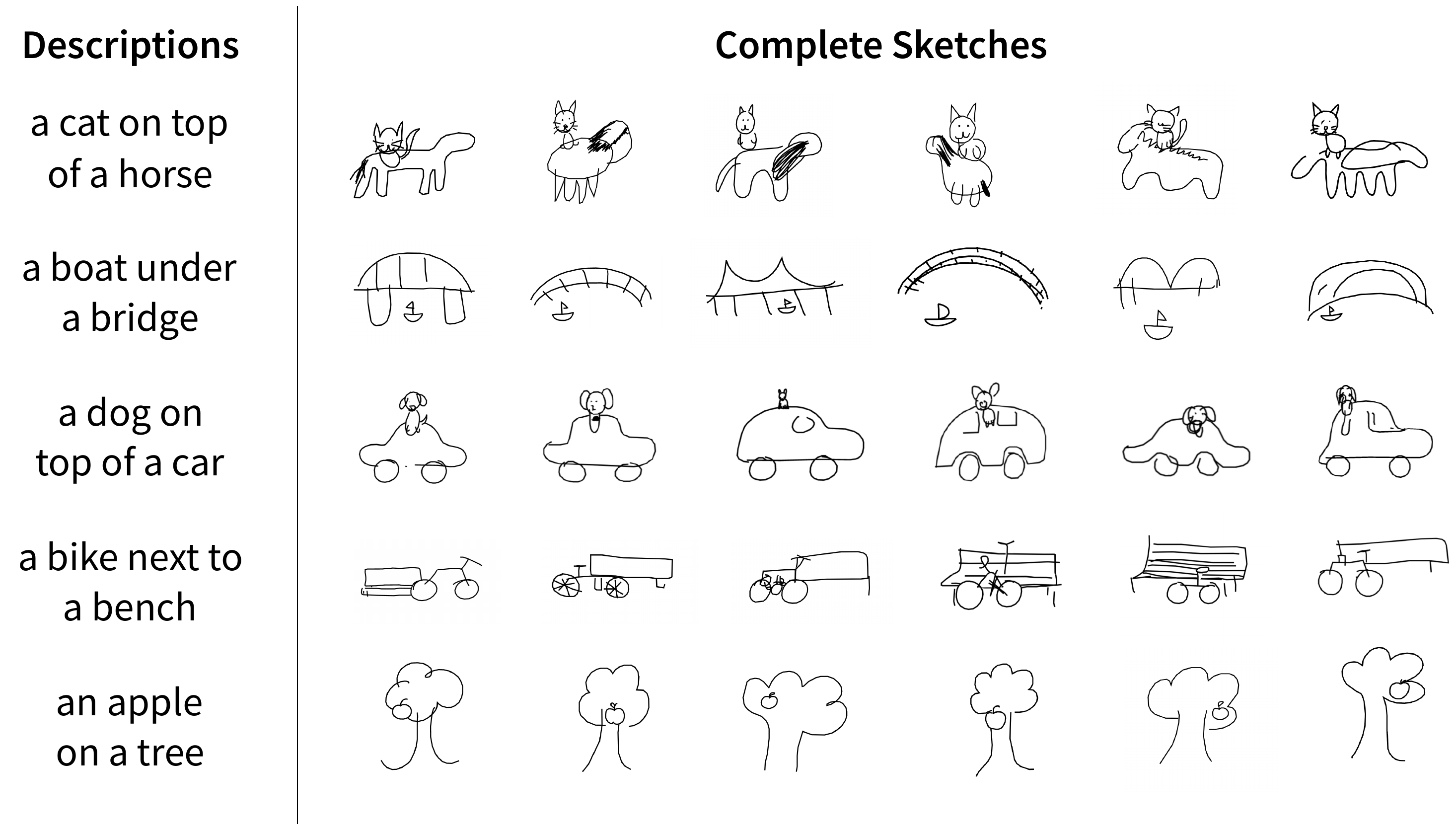}
  \caption{Complete scene sketches generated by \systemname. \systemname{} is able to generalize to novel concepts such as `a cat on top of the horse.'}
  \label{fig:complete_sketches}
\end{figure}

Combining the composition layouts and object sketch generation model of individual objects, \systemname{} generates complete scene sketches directly from text descriptions. Several examples of the sketches are shown in Figure~\ref{fig:teaser} and Figure~\ref{fig:complete_sketches}. In these figures, sketches that correspond to `a boat under a bridge' consist of small boats under bridges, whereas using `an apple on a tree' as the input creates sketches with small apples on large trees that follow the actual sizes and proportions of the objects. Moreover, \systemname{} is able to generalize to novel concepts of `a cat on top of horse,' such that the only relations involving a cat and a horse in the Visual Genome dataset which the model was trained on is `a horse facing a cat.' The sizes of cats and horses in these sketches are in proportion to their actual sizes and the cat is adequately placed on the back of the horse.

\subsection{Human perception user-study}
\systemname's high-level goal is to augment users' communication ability in sketches by generating realistic, plausible and coherent sketches for users to interact with and take reference from in their learning and communication processes. To complement the quantitative and qualitative evaluation of the sketches, we conducted user studies on Amazon Mechanical Turk (AMT) to gauge human subjects' opinions on the realism and ability of the sketches in conveying the description used to generate them.

\subsubsection{Study Procedure}
\begin{figure}[ht]
  \centering
  \includegraphics[width=0.9\linewidth]{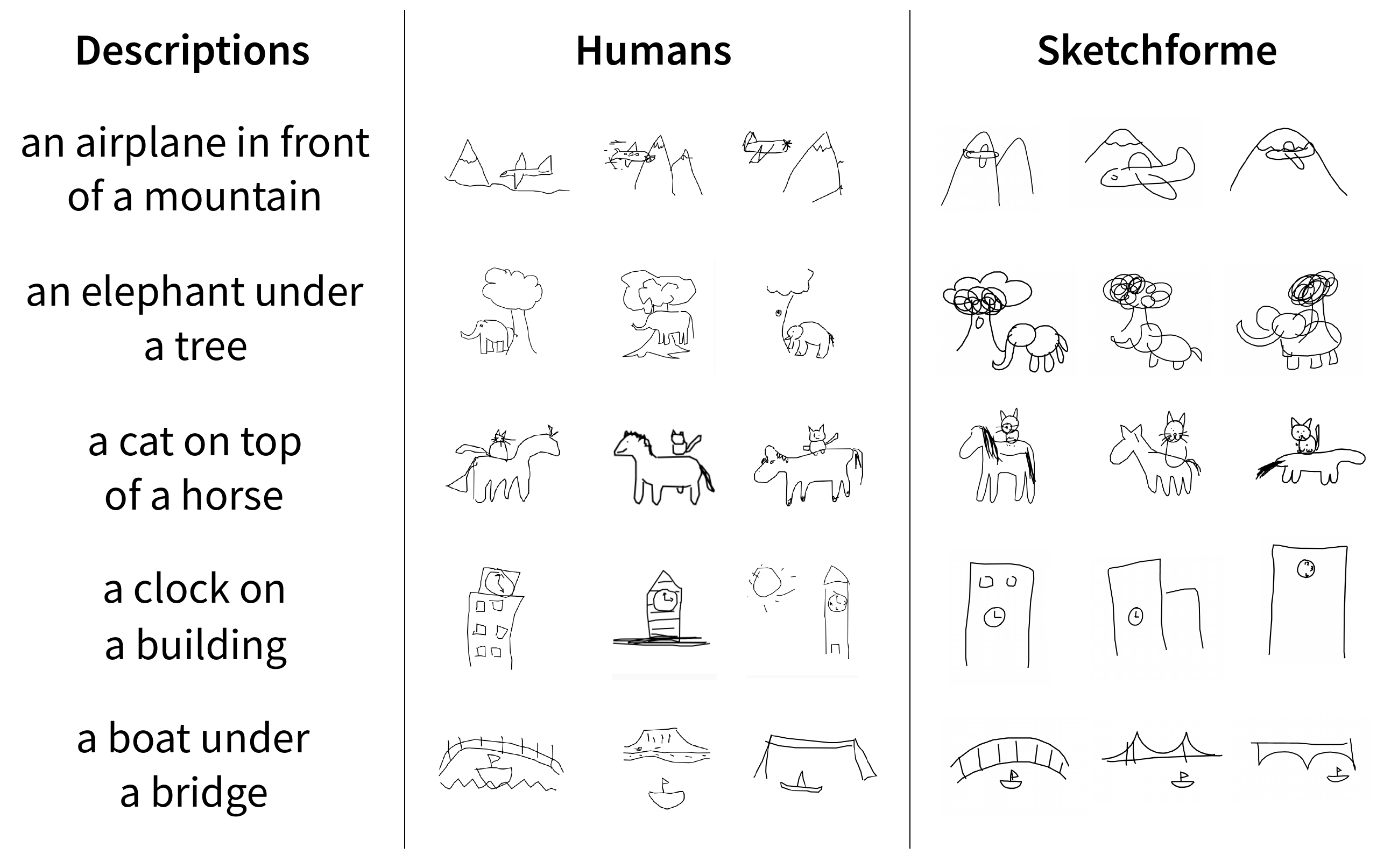}
  \caption{Samples of Sketches produced by humans and \systemname{} used in the AMT user study.}
  \label{fig:user_sketches}
\end{figure}

We recruited 51 human subjects on AMT and asked them to each review 50 sketches generated by either humans or \systemname{}. These 50 sketches are generated from five descriptions. The human-generated sketches are obtained from another AMT task prior to this user study based on Quick, Draw! \cite{quickdraw}. These human-generated sketches are shown in Figure \ref{fig:user_sketches}. In this study, subjects are provided with the complete sketched scene and the descriptions that the scenes are based on. Subjects are required to respond to the following questions: 

\begin{enumerate}
    \item{Do you think this sketch was generated by a computer (AI) or a human?}

    \item{On a scale of 1-5 (1 represents that description conveyed very poorly, 5 represents that description conveyed very well), how well did you think the message is conveyed by the sketch?}
\end{enumerate}

The subjects are given 10 sketches as trial questions with answers to (1) provided to them at the beginning of the task. After completing the trial tasks, the subjects' answers to the remaining 40 sketches are aggregated as the actual study result. This study protocol is similar to perception studies commonly used to evaluate synthetic visual content generation techniques in the deep learning community \cite{pix2pix}. In addition, we collected comments from the users (if any) and their perceived overall difficulty of the task at the end of the task.

\subsubsection{Results}
\begin{figure}[ht]
  \centering
  \includegraphics[width=1.0\linewidth]{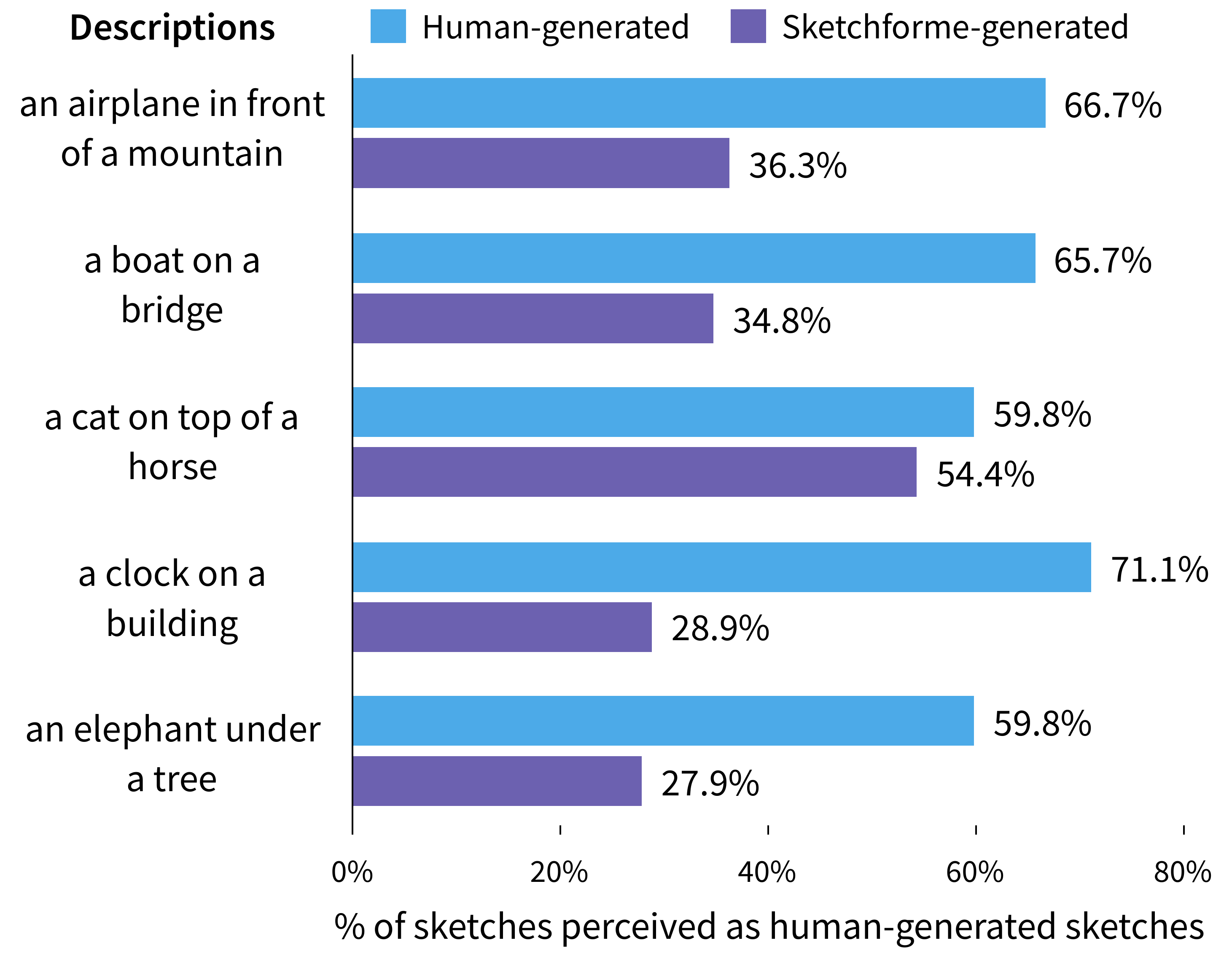}
  \caption{Percentage of sketches considered by users as human-generated. On average, 64.6\% of human-generated sketches are perceived as human-generated, while 36.5\% of \systemname-generated sketches are perceived as human-generated.}
  \label{fig:amt_q1}
\end{figure}

The first question probes the realism of the sketches with a Turing-test-style question asking the subjects to determine whether the sketches are created by humans. Subjects on average considered 64.6\% of the human-generated sketches as generated by humans, while they considered 36.5\% of \systemname-generated sketches as generated by humans as shown in Figure \ref{fig:amt_q1}. Although the percentage of \systemname{}-generated sketches considered as generated by humans are statistically significantly lower ($p = 1.05 \times 10^{-10}$, paired t-test) than human-generated sketches, individual participants commented in the study that it was difficult to distinguish between human-generated and \systemname-generated sketches. P2 mentioned that they "really couldn't tell the difference in most images." P6 commented that they "didn't know if it was human or a computer." These results demonstrate the potential for \systemname{} in generating realistic sketched scenes. 

We hypothesize one of the possible reasons for the lower percentage of \systemname-generated sketches to be considered as human-drawn is that the curves of the synthetic sketches are in general less jittery than human-drawn sketches. We suggest future work explore introducing stroke variation to generate more realistic sketches.

\begin{figure}[ht]
  \centering
  \includegraphics[width=1.0\linewidth]{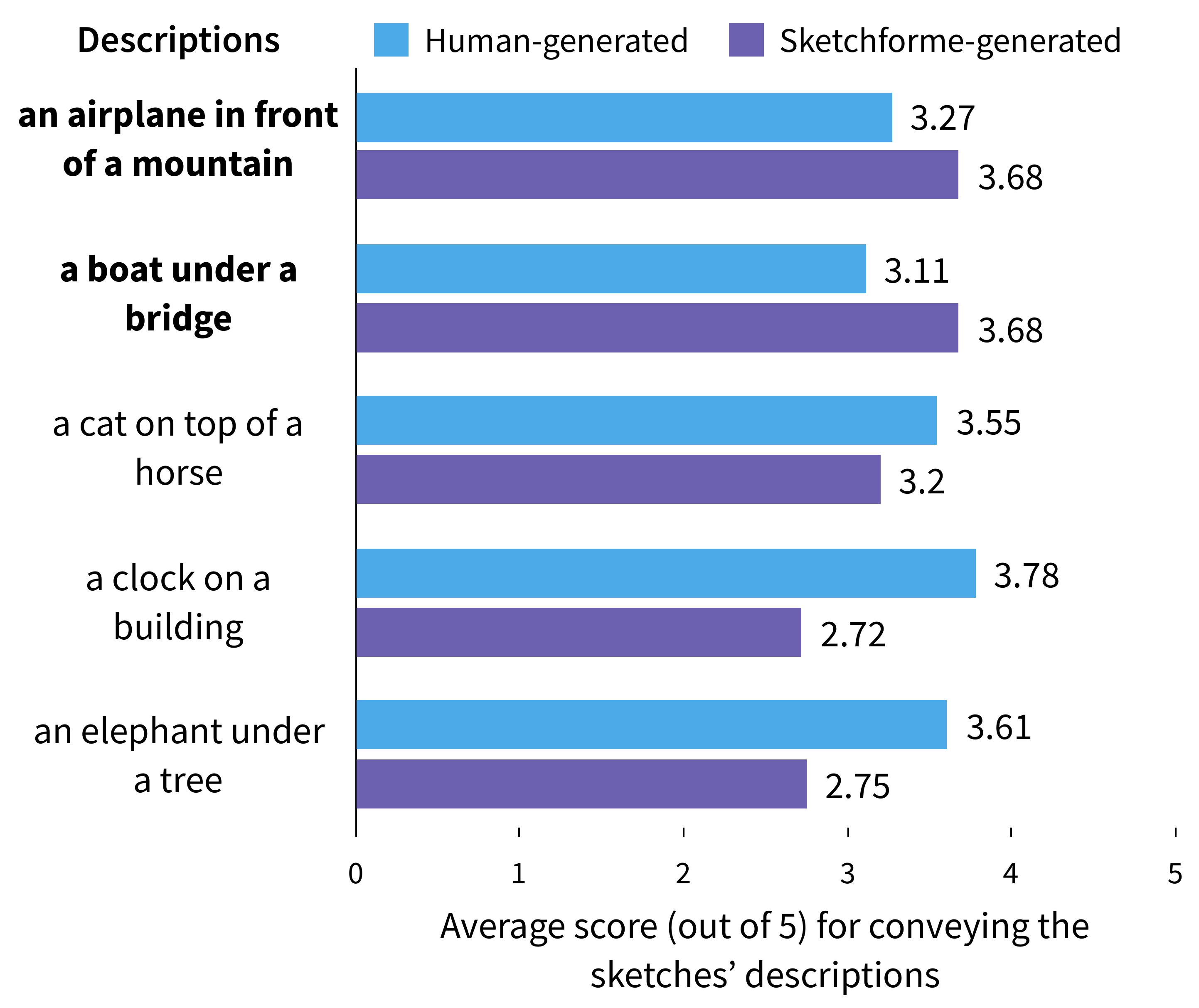}
  \caption{Average score for the conveying of descriptions by the sketches. \systemname-generated sketches perform better than human-generated sketches for sketches of `a boat under a bridge.' and `an airplane in front of a mountain.'}
  \label{fig:amt_q2}
\end{figure}

The results for the second question reflects the ability of the sketches to communicate the underlying descriptions of the sketches. The average score for human-generated sketches is $\mu = 3.46$, whereas the average score for \systemname-generated sketches is $\mu = 3.21$ as shown in Figure \ref{fig:amt_q2}. Although the \systemname-generated sketches achieved lower scores overall, \systemname-generated sketches achieved statistically better average scores for sketches based on two of the descriptions: `a boat under a bridge' and `an airplane in front of a mountain' ($p < 0.0005$, paired t-test). There is also no significant difference between the scores of human/\systemname{}-generated sketches based on `a cat on top of a horse'. This shows the competitive performance of \systemname-generated sketches in communicating underlying descriptions for some scenes.

\section{Applications}
In this section, we explore several applications that can benefited from \systemname{}'s ability to synthesize compelling sketches from natural language descriptions.

\label{sec:applications}
\subsection{Sketch-assisted Language Learning}
Sketches have been shown to improve memory \cite{psyc}. As language learning is a memory-intensive task, \systemname{} could support language education applications based on sketches. These sketches can potentially create engaging and effective learning processes and avoid rote learning. 

\subsubsection{Language Learning Application}
To explore the possibility of \systemname{} in supporting language learning, we built a basic language-learning application that aims to educate learners with a translation task from German to English. In this application, learners are presented with a German phrase, and are asked to translate it to English in the form of multiple choices similar to the process of learning term definitions from flash-cards. This application also implements the Leitner system \cite{leitner} with three bins that repeats phrases that learners make the most mistakes on most frequently. Using this system, the phrases are moved to different bins depending on the participants' familiarity of the translations.

\begin{figure}[ht]
  \centering
  \includegraphics[width=1.0\linewidth]{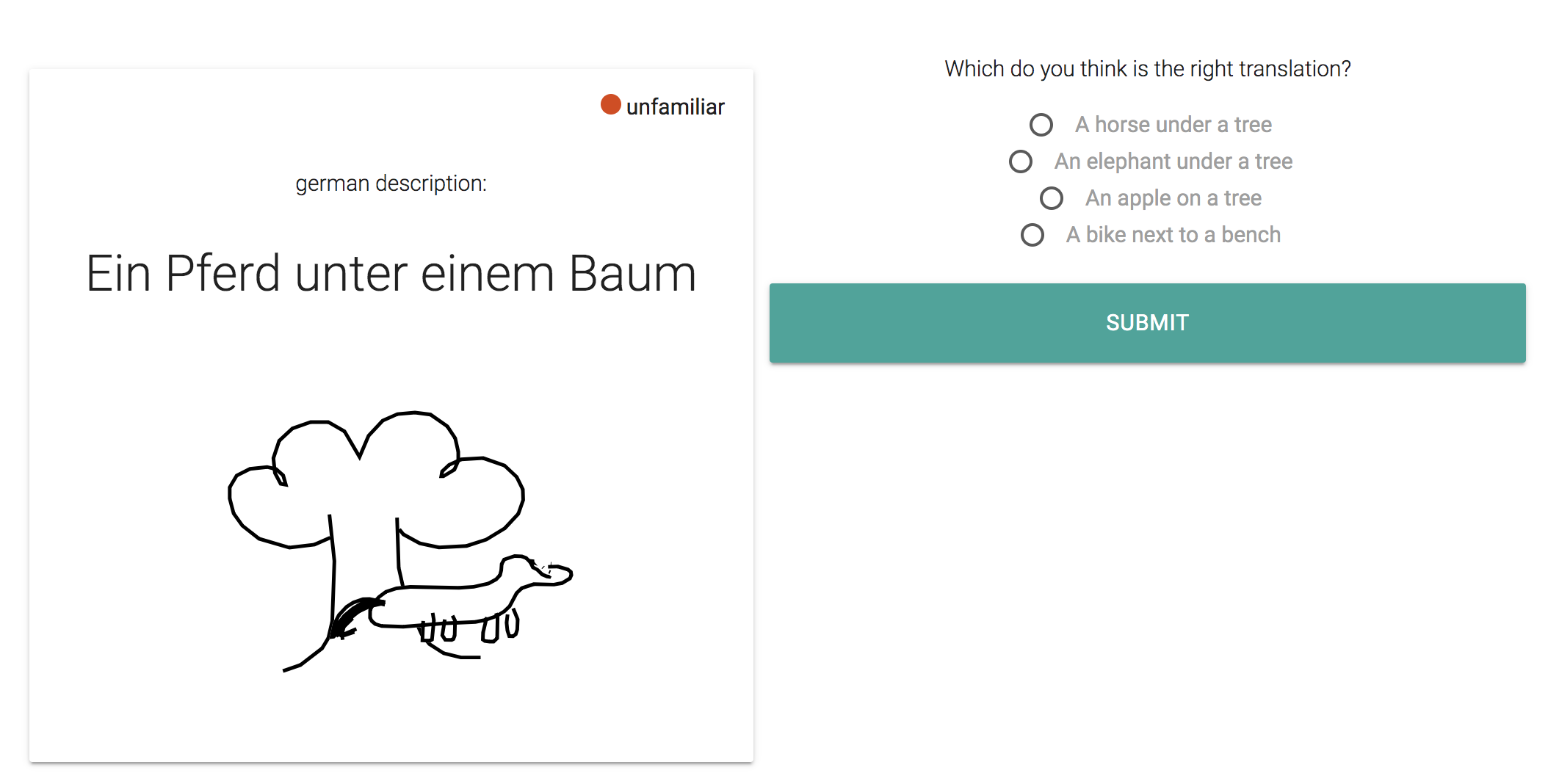}
  \caption{User interface for the language learning application powered by \systemname{}. Sketches significantly reduced the time taken to achieve similar learning outcome.}
  \label{fig:language_ui}
\end{figure}

We gathered 10 pairs of German-English sentences from a native German speaker and form 2 sets of 5 translations each. In addition, deceptive English sentences are added as other choices in the multiple-choice test to be selected by the learners in the application. We deployed this application on AMT to test the improvement on learning performance by presenting \systemname{}-generated sketches along with the phrases. The full application with the sketches presented to the users is shown in Figure \ref{fig:language_ui}.

\subsubsection{Study Procedure}
The study consists of a training phase and a testing phase for each participant. In the training phase, participants are presented with correct answers after answering each question. The participant can only advance to the next phase when they answer all questions correctly consecutively for all translations according to the Leitner system. In the testing phase, participants are given one chance to provide their answer to all translations without seeing the correct answers. The participants are divided into two conditions, with the 'control' group only receiving phrases on their interface during training, and the 'treatment' group that receives both phrases and sketches generated by \systemname{} on their interface during the training phase. Both groups receive only the phrases on their interface during the testing phase. Moreover, we use our two sets of translations for training and testing phases alternatively, such that the participants will not get consecutive training and testing phases for the same set of descriptions. 

The performance of the participants during the study are monitored with multiple analytical metrics including completion time of each phases and scores on the test phase etc. At the end of the study, we also provide surveys for them to rate the difficulty of the task and the usefulness of the sketches (if applicable) on five-point Likert scales, and ask them to provide any additional suggestions to the interface. 

\subsubsection{Results}
\begin{figure}[ht]
  \centering
  \includegraphics[width=0.8\linewidth]{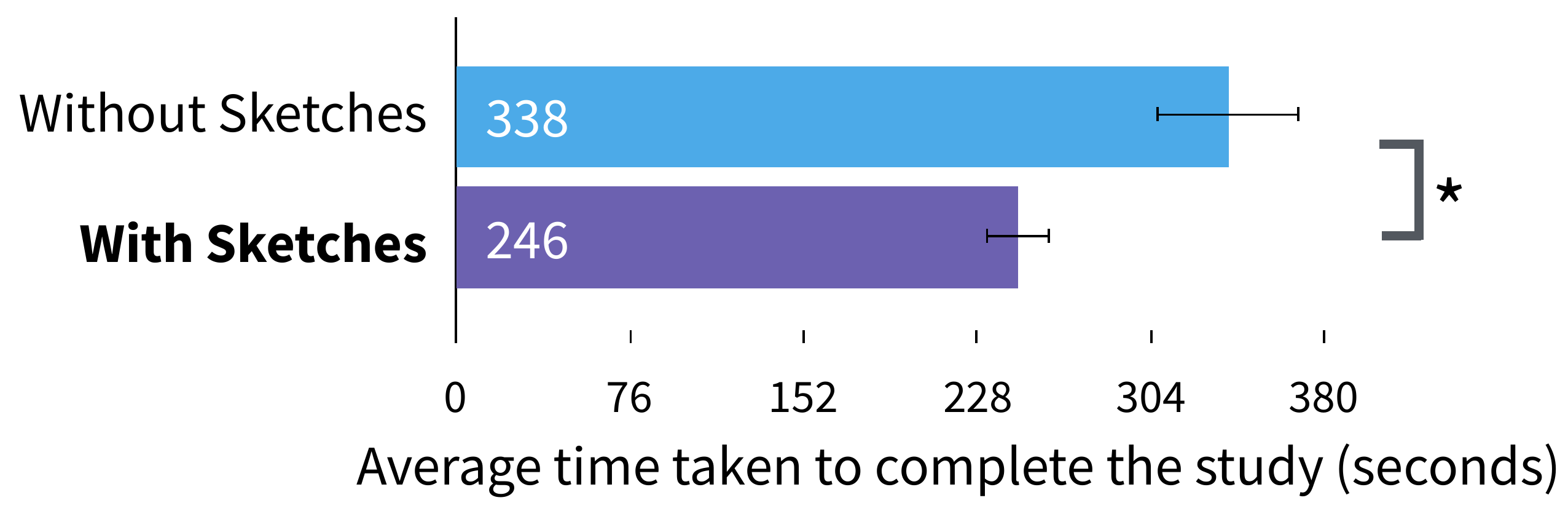}
  \caption{Average time taken to complete the language learning task among groups. With \systemname-generated sketches, participants take significantly less time on completing the study.}
  \label{fig:lang_app}
\end{figure}

We recruited 38 participants on AMT to participate in the study. While we did not see significant differences ($p = 0.132$, unpaired t-test) in the correctness of answers in the testing phase of the phrases between the `control' and `treatment' groups of participant, we discovered that the \emph{time taken to complete the learning task} was significant less ($p = 0.011$, unpaired t-test) for the `treatment' group at $\mathbf{246}$ seconds on average, compared to $\mathbf{338}$ seconds for the control group as shown in Figure \ref{fig:lang_app} . Moreover, from the post-study survey, we also discovered that the `treatment' group in general found the sketches to be helpful for learning (rated $4.58$).

As \systemname{} is an automated system that is capable of generating sketches from free-form text descriptions, and with these promising results on sketch-assisted language learning, we envision \systemname{} to support and improve large-scale language learning applications in the future. 

\subsection{Intelligent Sketching Assistant}
We designed \systemname{} to support interactive sketching systems using a sequential architecture and a multi-step generation process. To demonstrate \systemname{}'s capability of supporting interactive human-in-the-loop sketching applications, we built a prototype of an intelligent sketching assistant reflective of two potential use-cases:  

\subsubsection{Auto-completion of scenes}
As \systemname{}'s Scene Composer is a sequential architecture that takes the previous object in the scene to generate the next object, we can complete unfinished user scenes instead of starting with a blank canvas by starting the generation with both the start token $b_0$ and an existing object in the scene created by the user $b_1$. Figure \ref{fig:choices} shows examples of \systemname{} completing users' sketch of a horse in step a by adding the other object involved in the scene.

\subsubsection{User-Steerable generation}
\begin{figure}[ht]
  \centering
  \includegraphics[width=0.8\linewidth]{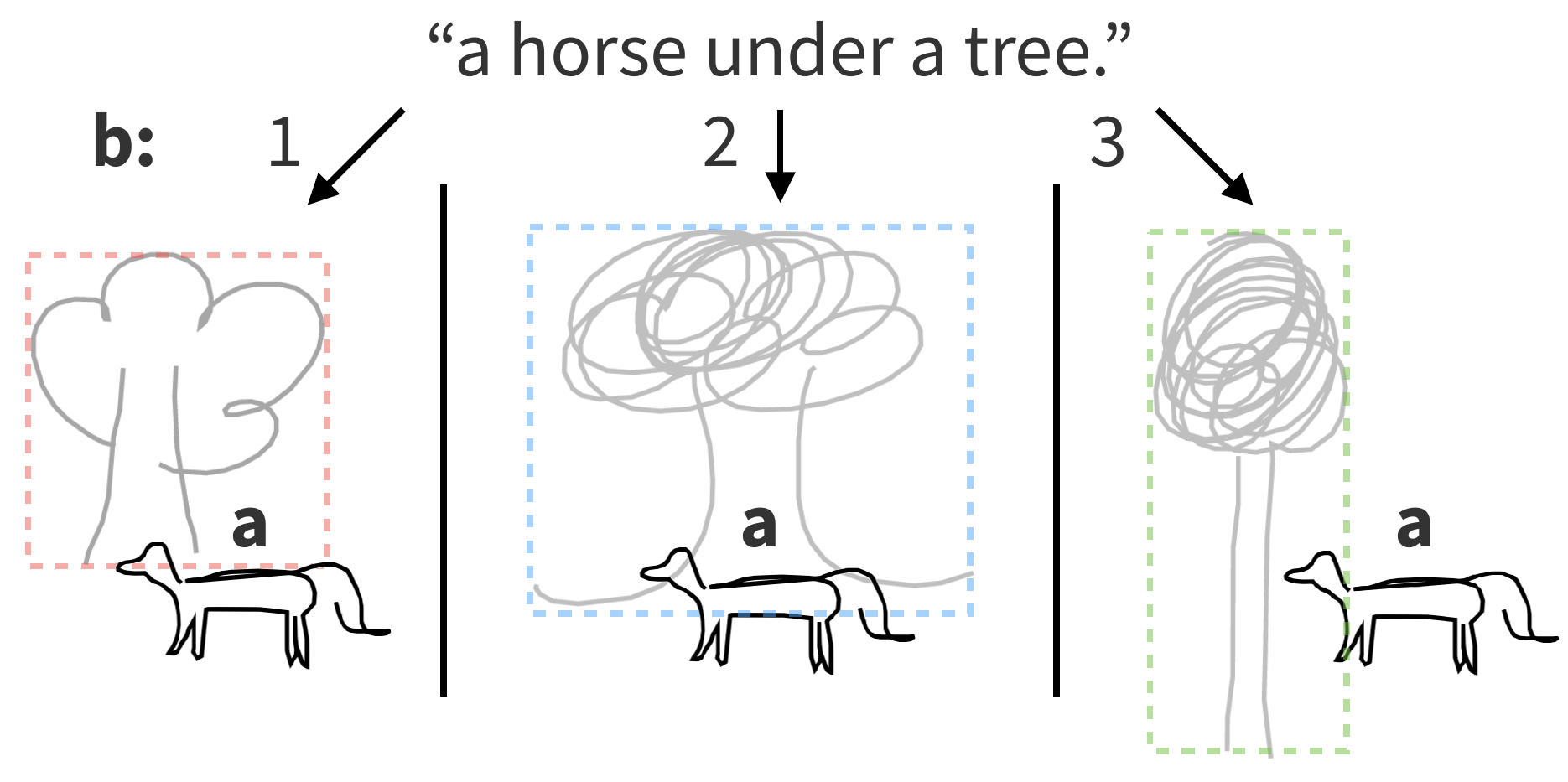}
  \caption{Intelligent sketching assistant powered by \systemname{}. The user can a) create partial sketch for \systemname{} suggest multiple candidates for b) users to choose to adequate sketch they prefer from the description `a horse under a tree.'}
  \label{fig:choices}
\end{figure}

\systemname{}'s Scene Composer is capable of generating multiple potential candidate objects at each step while composing the scene layout of the generated sketch. As such, users can select their preferred scene layout from multiple potential candidates. Figure \ref{fig:choices} shows multiple candidates proposed by \systemname{} based on a text description in step b. Moreover, since Sketch-RNN is also capable of generating a variety of sketches, the users can also select their preferred sketches of each individual objects in the scene.

\section{Limitations}
\subsection{Occlusions and layer order}
\systemname{} is trained to encode composition principles from a natural image dataset. In natural images, objects might occlude each other, hence affecting the sizes and positions of the bounding boxes in the composition layouts. Figure \ref{fig:poses}a shows several boats that was inadequately placed in front of parts of the bridge that should have occluded the boats. To overcome this limitation, future systems can augment \systemname{} by including advanced vision models to determine the layer order in the original natural image. The current \systemname{} system only considers a naive layer order determined by the sequence of generation of the composition layouts.

Moreover, having occluded compositions lead to the generation of overlapping sketches requiring additional research on realistic methods to handle overlapping sketched objects. For instance, the model that generates composition layouts can enforce constraints to avoid overlaps in the sketches, or follow hand-crafted rules to handle overlaps.

\subsection{Aspect ratios might be weak signals for object poses}
\begin{figure}[h]
\begin{center}
\subfloat[Occluded Objects]{\includegraphics[width=0.4\linewidth]{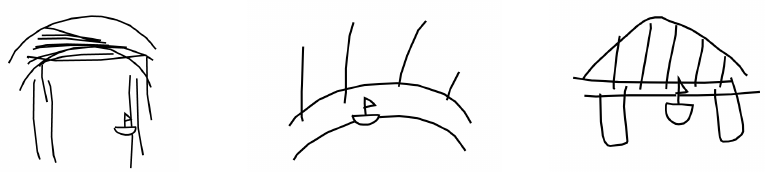}}
\qquad
\subfloat[Incoherent Poses]{\includegraphics[width=0.4\linewidth]{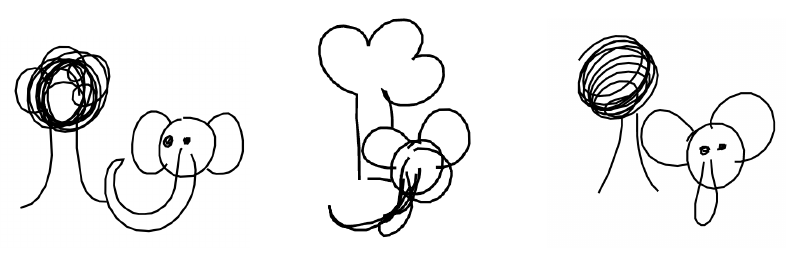}} 
\end{center}
\caption{Limitations of \systemname's sketching process. In a), the boat is significantly occluded by the bridge which affects the quality of generated sketches. In b), the elephant was provided with a square bounding box. This guides the system to sketch only the face of the elephant which is inappropriate for the scene.}
\label{fig:poses}
\end{figure}

\systemname{} utilizes the aspect ratios of bounding boxes as the primary signal to inform the shapes of sketches of individual objects. These shapes may consequentially determine the poses of sketched objects. Although these shapes can be sufficient to determine correct poses for some object classes, such as the `tree' class, constraining the shapes might be weak signals for other object classes. These shapes can suggest incoherent perspectives or partial sketches such as examples shown in Figure \ref{fig:poses}b. In Figure \ref{fig:poses}b, only the faces of the elephants were sketched due to the aspect ratios provided to the extended Sketch-RNN model, which is inappropriate for composing sketched scenes. To mitigate this limitation, future work should model the pose of objects in sketches and natural images to augment the composition knowledge of their models, such as incorporating complete masks of the objects.

\section{Future Work}
\label{sec:future-work}
\subsection{Conversational sketch suggestion and tutorial system}
With the highly interactive media of sketching and language, combined with \systemname's high-level and low-level understanding of each element of the sketched scene, we believe future work should explore conversational interfaces for generating sketches and interactive tutorial systems that guide users to create sketches coherent to text descriptions. Moreover, since the Object Sketcher in \systemname's generation process is capable of completing partial sketches created by users, \systemname{} can suggest possible strokes following incomplete user sketches at the object level, which can be useful in sketch education applications.

\subsection{Educational applications in other domains}
The unique interplay between natural language and sketches embodied by \systemname{} creates possibilities of building new applications that utilize the interactive properties of sketching and language. In this paper, we explored the capability of \systemname{} in supporting basic language learning. We believe future work could explore other domains such as science and engineering education as text-annotated sketches are frequently used in these domains.

\subsection{Coloring and animation}
The current sketches generated by \systemname{} are binary sketch strokes without colors or animations. Future work should explore colored and/or animated sketches to enable richer user experiences. For instance, the natural image datasets that \systemname{} used to train models in the first step of sketch-generation process can be used to determine possible colors of the sketched objects.

\section{Conclusions}

This paper presents \systemname, a novel sketching system capable of generating abstract scene sketches that involve multiple objects based on natural language descriptions. \systemname{} adopts a novel two-step neural-network-based approach: the \emph{Scene Composer} obtains \emph{high-level} understanding of layouts of sketched scenes, and the \emph{Object Sketcher} obtains \emph{low-level} understanding of sketching individual objects. These models can be trained without text-annotated datasets of sketched scenes. In the user study evaluating the expressiveness and realism of sketches generated by \systemname, human subjects considered \systemname-generated sketches more expressive than human-generated sketches for two of the five seeded descriptions. They also considered 36.5\% of these sketches to be generated by humans. The sketches generated by \systemname{} significantly improved a sketch-assisted language learning system and enabled compelling intelligent features of a sketching assistant. \systemname{} possesses the potential to support interactive applications that utilize both sketches and natural language as interaction media, and afford large-scale applications in sketching, language education and beyond.

\balance{}

\bibliographystyle{SIGCHI-Reference-Format}
\bibliography{main}

\end{document}